# How well should the DENIS Survey probe through the Galactic Plane?

Gary A. MAMON

*Institut d'Astrophysique, 98 bis Blvd Arago, F–75014 Paris, FRANCE*
*& DAEC, Observatoire de Paris-Meudon, F–92195 Meudon, FRANCE*

**Abstract.** The DENIS 2 micron survey is described. It's ability to probe galaxies at very low galactic latitudes is computed using simulated images. This analysis indicates that DENIS will be able to separate galaxies from stars, right through the Galactic Plane, for galactic longitudes $45° \leq \ell \leq 315°$, with a maximum loss in the magnitude limit for reliable extraction of $\simeq 1.2$ magnitude relative to that at the Galactic Pole (where the 92% reliability limit is $K' = 13.2$ for star/elliptical galaxy separation, and fainter for other morphological types). Extinction corrected magnitude limits are substantially worse, but for $|b| \geq 2°$, they are within one magnitude from that of the Galactic Poles. Thus, confusion by stars will not prevent the establishment of a reliable galaxy catalog right through the Galactic Plane. However, extinction will produce galaxies of lower surface brightness, which will be more difficult to detect.

## 1. Introduction

The recent advent of large 2D detectors, sensitive to light in near-infrared wavelength bands, holds great promise for extragalactic astronomy. Indeed, our view of the Universe is hampered by the dust layer in the Galactic Plane: in optical wavebands this dust is optically thick, and the optical detection of galaxies is severely reduced at galactic latitudes $|b| < 5°$ (*e.g.*, Weinberger, Woudt, Kraan-Korteweg, all in these proceedings). In the far infrared, probed by the IRAS satellite, this cool dust emits thermally and causes confusion with external galaxies, which becomes serious for $|b| < 12°$ within $90°$ from the Galactic Center (Meurs & Harmon 1988; see also Meurs and Saunders, both in these proceedings).

Similarly, our view of external galaxies is hampered by extinction from their interstellar dust in the optical, and thermal emission by this dust in the far infrared.

Moreover, both the optical (especially blue) and far-infrared luminosities of galaxies are enhanced by recent star formation, and it is thus believed that the near infrared light correlates best with the stellar content of galaxies, independent of morphological type (see Jablonka & Arimoto 1992).

With these advantages in mind, various members in the extragalactic community in Europe joined DENIS (*DEep Near-Infrared Southern Sky Survey*), while many of our American colleagues joined the very similar 2MASS (*2 Mi-* *cron All Sky Survey*) project (see Huchra, in these proceedings). These two surveys should provide detailed mapping of the local Universe, without being biased in favor of star forming galaxies. The applications of near-IR surveys to the study of groups and clusters of galaxies have been reviewed by Mamon (1994). The aim of the present work is to check how well will the DENIS survey probe large-scale structure of the Universe behind the Galactic Plane.

## 2. Overview of DENIS

The DENIS survey will map the entire southern sky ($-90° \leq \delta \leq +2°$) in the $I$ (0.8$\mu$m), $J$ (1.2$\mu$m), and $K'$ (2.1$\mu$m) wavebands. The observations will make use of the existing ESO 1m telescope, on a 2/3 time basis until August 1994, and in principle full-time after that. The survey is expected to begin around the end of 1994 and last 3 years. Rapid mapping imposes a large field of view (12′), which amounts to 3″ pixels for the $J$ and $K'$ bands, using the current state-of-the-art 256 × 256 HgCdTe NICMOS-3 arrays designed by Rockwell, and 1″ pixels in the $I$ band using a Tektronix 1024 × 1024 CCD.

The sky will be scanned in declination along strips, 30° long and 12′ wide, in a stop-and-stare mode. Elementary images of 12′ × 12′ are integrated for 9 × 1 s in $J$ and $K'$, making use of a micro-scanning device to dither the images by 1/3 pixel, yielding a pseudo-resolution in $J$ and $K'$ of 1″.

Preliminary calculations indicate that the DENIS survey will detect roughly $10^8$ stars ($K' < 14.5$) and 250 000 galaxies ($K' < 13.7$), the latter with reliable star/galaxy separation and $K'$-band photometry accurate to 0.2 magnitudes (Harmon & Mamon 1993).

The raw data will be reduced at the Paris Data Analysis Center (PDAC), then shipped to the Leiden Data Analysis Center (LDAC) for the extraction of point sources and the bright galaxies, while the faint galaxies will be extracted back at the PDAC.

## 3. New image simulations

The early image simulations presented by Harmon & Mamon (1993) took into account neither the range of morphological types of galaxies, nor the confusion with stars and extinction at low galactic latitudes. These issues are addressed here.

Artificial images are simulated by placing galaxies, of a given morphological type (*Ellipticals*, *Edge-On Disks*, and *Face-On Disks*), on a square grid, where in each row, galaxies have the same total magnitude. Stars are overlaid at random positions, according to a home-built star-count program, similar in spirit to that of Bahcall & Soneira (1980). An image is built using the IRAF ARTDATA package, taking into account the parameters of the atmosphere, telescope, and detector. A Moffatt PSF ($\beta = 2.5$) is assumed.

For the $J$ and $K'$ bands, the 9 subimages for each exposure are interlaced into a single image, with 768 × 768 1″ pseudo-pixels, with the information on 1 pseudo-pixel taken from that of *the* concentric 3″ pixel, which comes from a single subimage. In fact, images twice the size of the DENIS images (1536 × 1536 pixels of 1″) are used, for increased statistics (grids of 20 × 20 galaxies). The image is saved in 16-bit format as in the survey, with the same sampling ($\sigma = 2\,\mathrm{ADU}$).

The input stellar luminosity functions in the $I$ and $J$ bands are taken from Mamon & Soneira (1982). The $K$-band stellar luminosity function is adapted from Wainscoat et al. (1992), which appears to provide significantly better fits to existing star count data than the $K$ luminosity function of Mamon & Soneira. Its parameters, in the notation of Bahcall & Soneira and Mamon & Soneira, are $\alpha = 0.40$, $\beta = -0.01$, $1/\delta = 1.0$, $M* = 3.30$, and $n* = 0.014\,\mathrm{mag}^{-1}\,\mathrm{pc}^{-3}$. The random stars are placed down to a magnitude limit 1 magnitude fainter than the theoretical magnitude limit for $5\,\sigma$ detection of point sources, so that the faintest stars contribute to increased noise in the diffuse background.

Rather than assume, as Bahcall & Soneira (1980), a plane parallel absorbing sheet, we adopt an absorbing layer that has an exponential distribution both in height (scale-height = 100 pc) *and* in length (scale-length equal to the stellar scale-length of 3.5 kpc). We fix $A_V = 0.15$ to infinity at the Galactic Pole and $A_V = 30$ from the Sun to the Galactic Center (assumed to be 8 kpc distant). Because the absorbing layer is very heterogeneous, we will compare, in the future, with the Burstein & Heiles (1982) maps of galactic extinction for $|b| > 10°$.

The bulge (or spheroid) of our Galaxy is assumed to follow Young's (1976) deprojection of the $r^{1/4}$ law, with axis ratio 0.9, and effective radius of 8 kpc. The local disk to bulge number density ratio is chosen as 800.

Elliptical galaxies are modeled as $r^{1/4}$ laws, while disk galaxies are exponential disks. The extrapolated central unattenuated blue surface magnitude is $\mu_B(0) = 14.0$ for ellipticals and 21.3 for disks. After determining the half-light radius, the galaxies are attenuated by the extinction to infinity returned by the star count program.

Table 1 below lists some of the DENIS parameters used in these simulations

Table 1.  Input Parameters

|  | $I$ | $J$ | $K'$ |
|---|---|---|---|
| $\lambda\,(\mu)$ | 0.8 | 1.2 | 2.1 |
| $A_\lambda/A_V$ | 0.48 | 0.28 | 0.11 |
| $\mu_{\mathrm{sky}}$ (mag arcsec$^{-2}$) | 19.0 | 16.0 | 12.5 |
| Seeing (″ FWHM) | 1.2 | 1.1 | 1.0 |
| Quantum efficiency | 0.60 | 0.60 | 0.65 |
| Optical transmission | 0.35 | 0.30 | 0.30 |
| Read-out noise ($e^-$) | 35 | 50 | 50 |
| Integration time (s) | 7.5 | 9 × 1 | 9 × 1 |
| Pixel size (″) | 1 | 3 | 3 |
| Disk color ($B - C$) | 1.4 | 1.9 | 2.8 |
| Bulge color ($B - C$) | 2.5 | 3.2 | 4.2 |

$B - C$ is the color $B - I$, $B - J$, $B - K'$, for bands $I$, $J$, and $K'$, respectively.

## 4. Galaxy Extraction

There are various steps in extracting galaxies from images:

### 4.1. Detection

One wants to have a *complete* sample, in terms of a high probability that given an object, it is detected. One wants to have a *reliable* sample, so that a detected object is not caused by noise. Recent work of ours (Mamon & Contensou 1993) has convinced us that the interlaced images must be *smoothed* to detect galaxies, which are low surface brightness objects, in comparison with stars. The smoothing filter could be a boxcar, gaussian, or something else. It turns out that there is little difference in performance between boxcar and gaussian filters *in high galactic latitude fields*, but we have not yet tested the differences in crowded fields, for which galaxies will often be blended with stars. This is an important issue because in low star density regions, the smoothing scales need to be roughly 3 to 4 times greater to optimally extract the face-on disk galaxies, than to optimally extract the edge-on galaxies, the ellipticals, and the stars themselves. We hope to gain by first removing the stars detected by the LDAC. Note that a $3 \times 3$ boxcar smoothing amounts to transforming the interlaced image into an image where each $1''$ pixel corresponds to the average of the 9 $3''$ pixels containing that $1''$ pixel (one per subimage).

### 4.2. Star-Galaxy Separation

We wish to have a reliable star/galaxy classifier, especially in the sense that objects called galaxies will not in fact be stars. This is complicated by the fact that the ratio of star to galaxy number densities ($K' < 14$) varies from 10 to $> 10^4$ from the Galactic Poles to the Galactic Plane.

A simple star-galaxy separation algorithm is applied to each object in the *unsmoothed* image. The algorithm computes the mean surface brightness above a threshold for pixels within a given circular aperture around the true center of the object. This statistic should discriminate well between galaxies and blended stars. The aperture is set to be small, to avoid blending in with neighboring stars, although not too small to avoid noisy statistics. The threshold is a few background standard deviations above the background.

### 4.3. Photometry

For most cosmological applications, we wish to have a galaxy sample with accurate photometry ($\Delta m \leq 0.2$). We have performed tests of isophotal photometry, which show that the mean offset between isophotal magnitude and total magnitude depends on the isophote, the type of the object, and sometimes its magnitude. Presumably, this mean offset can be modeled. The photometric accuracy will be the combination of the dispersion of the offset and the uncertainty on its mean. These issues are still under investigation. In crowded fields, photometry will be biased by neighboring stars. The hope again is to gain by first removing the stars detected by the LDAC.

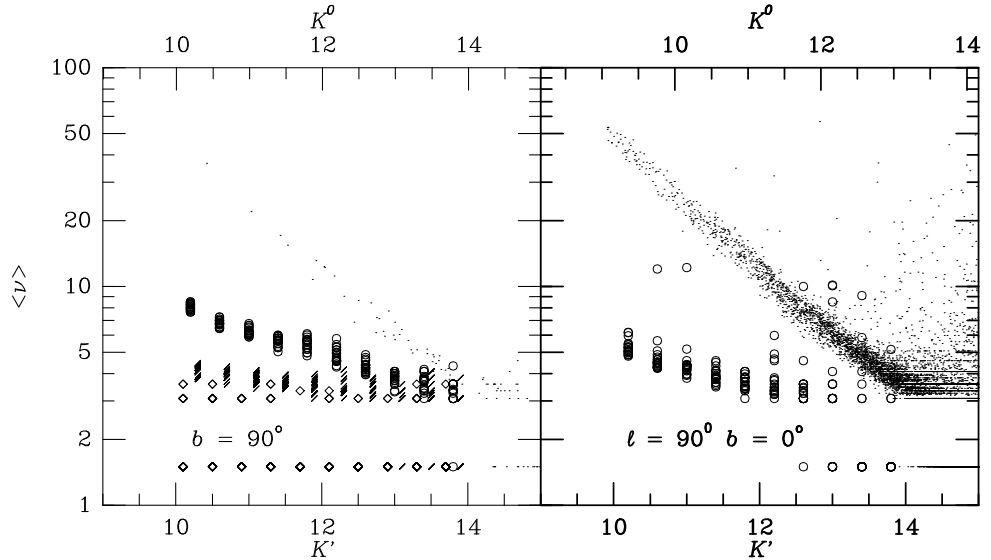

Figure 1. Star galaxy separation in the $K'$ band: mean surface brightness above threshold versus magnitude. Ellipticals, edge-on disks, face-on disks, and stars are shown as *circles*, *dashes*, *diamonds*, and *points*, respectively. The upper scale is the equivalent $K'$ magnitude for elliptical galaxies, after correction for extinction. The symbols at $\langle \nu \rangle = 1.5$ are objects for which no pixels were above the threshold.

## 5. Results on Star/Galaxy Separation

Figures 1, 2, and 3 plot the star/galaxy separation statistic (in units of the true background standard deviation) as a function of true object magnitude for the Galactic Poles (left-hand plots) and for $\ell = 90°, b = 0°$ (right-hand plots). Recall that the stars are field stars with the predicted number in each magnitude bin, while the galaxies are test galaxies (at an expected 20 galaxies per square degree at $K' < 14$, one expects 4 galaxies by quadruple-area DENIS like field, hence galaxy-galaxy confusion will be negligible). The galaxies that are above the galaxy locus are blended with stars, and the ordinate of such a galaxy indicates the surface brightness, hence magnitude of the star with which it is blended. The objects are assumed to have already been detected, and that their astrometry and photometry are relatively accurate. In the next step (in preparation), we plan to relax these assumptions by doing a global pass at detection, star/galaxy separation, astrometry and photometry.

The figures indicate that stars are more concentrated than galaxies, especially at bright magnitudes. At bright magnitudes, the mean surface brightness of stars scales as $\mathrm{Cst} + 2.1\,m$, where $m$ is the magnitude (one expects a slope of 2.5 at even brighter magnitudes). The limit for 92% reliable star/galaxy separation is $K' \simeq 13.2$ (see also Harmon & Mamon 1993) for ellipticals. The limits in $J$ are slightly better, thanks to the lower background, while in $I$, the limits are substantially better thanks to the increased angular resolution.

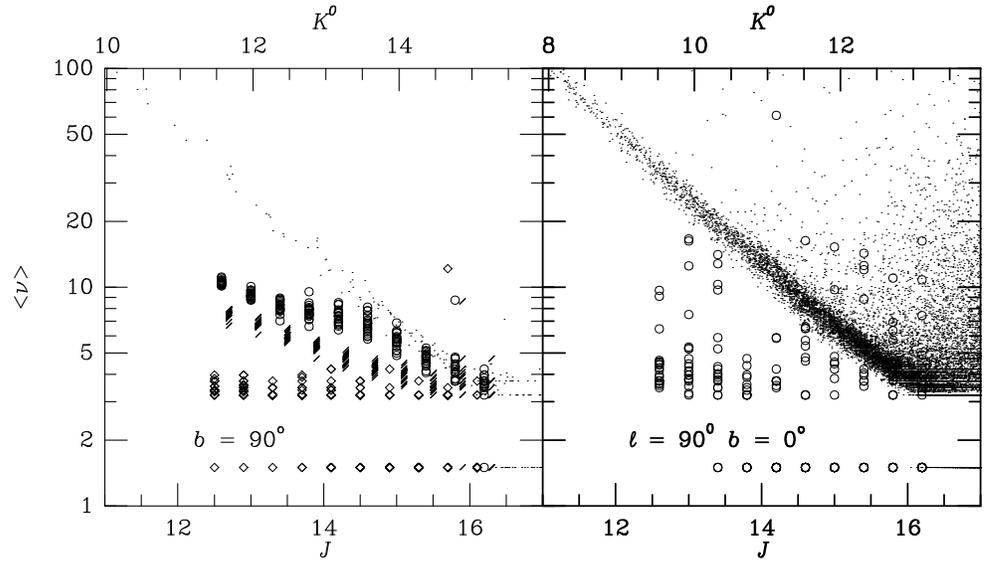

Figure 2.  Same as Fig. 1, but in the $J$ band.

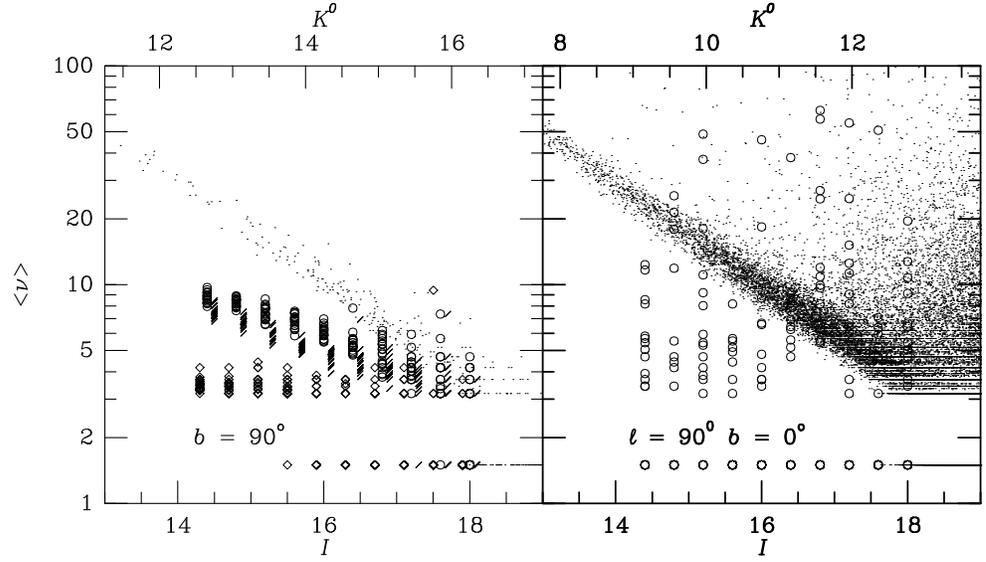

Figure 3.  Same as Fig. 1, but in the $I$ band.

The elliptical galaxies are more concentrated than the edge-on disk galaxies, which themselves are more concentrated than face-on disk galaxies. The 92% reliability limits are $K' \simeq 13.6$ for edge-on disks, and $K' \simeq 13.7$ for face-on disks. However, less than half of these latter-type galaxies will be detected at these magnitudes. Although they are hardest to detect, face-on disk galaxies, once detected, are the easiest to tell apart from stars.

The right-hand plots in Figures 1, 2, and 3 show the effects of confusion and extinction in the Galactic Plane. The stellar locus is unaffected by extinction (it just widens with increased stellar density), *i.e.*, the mean surface brightness of unresolved point-sources is a function of their magnitude and the point-spread-function. On the other hand, with higher extinction, the mean surface brightness of galaxies tends to decrease as d$\exp(-0.4\,A_\lambda)$. So, the galaxy locus is, in fact, better separated from the stellar locus in high extinction fields (but, for the same reason, galaxies will be more difficult to detect). However, in the $K'$ and $I$ bands, star/galaxy confusion brightens substantially the 92% reliability limiting magnitudes. This is much less so the case in the $J$ band, for which confusion always sets in at about the same magnitude as where the galaxy locus runs into the stellar locus for $|b| = 90°$ (see Table 2 below).

Table 2 provides a summary of the results of star/galaxy separation with the simple algorithm described above. One sees that, over most of the sky ($|b| \gtrsim 6°$), the $I$ band, thanks to its 3-fold increased angular resolution, provides superior star/galaxy separation, such that it will be able to separate galaxies that would not be detectable in $K'$ from stars. At faint magnitudes, the $I$ band is read-out noise limited. If the read-out noise is chosen to be $50\,e^-$ instead of 35, then the magnitude limits, in Table 2, should be decreased by about $2.5\log_{10}(50/35) \simeq 0.4$ magnitudes, and the $I$ band provides the best star/galaxy separation at high galactic latitudes for a read-out noise $\lesssim 120\,e^-$. In the $K'$ band, galaxies should be reliably separated from stars, *through the Galactic Plane*, with a loss of only 0.6 (1.2) magnitudes at $\ell = 90°$ (45°). However only few galaxies will be present up to these magnitudes, because of the strong dust extinction, even in $K'$. The corresponding losses in 92% reliability limiting magnitude, corrected for extinction, right through the Galactic Plane, are 1.4 and 3.6 magnitudes at $\ell = 90°$ and 45°, respectively, but only 1 magnitude or less for $|b| = 2°$ at these same longitudes.

Of course, the goal is not to have a highly reliable galaxy catalog everywhere, but instead a highly reliable galaxy catalog at high galactic latitudes and a moderately reliable (at the 50% level, see Table 2) galaxy catalog at low galactic latitudes. The results in Table 2 indicate that star/galaxy separation in the $K'$ band will have at least 50% reliability everywhere, except in a narrow region extending from $\ell = 0°$, $b = 3°$ to $\ell = 45°$ (and $\ell = 315°$), $b = 0°$, and roughly 92% reliability at the Galactic Poles. It remains to be seen how extinction affects the detection of low galactic latitude galaxies.

## 6. Prospects

In practice, the DENIS team will continue algorithmic development using simulated images, for the simple reason that we know everything we put in these images. In a second stage, we will take a dozen or so exposures of the same strips

Table 2. Reliability Limits for Star/Elliptical Galaxy Separation

| $\ell$ | $|b|$ | Reliability | $K'$ | $K^0$ | $J$ | $K^0$ | $I$ | $K^0$ |
|---|---|---|---|---|---|---|---|---|
| | 90° | 92% | 13.2 | 13.2 | 14.4 | 13.4 | 16.5 | 14.7 |
| | 90° | 50% | 14.2 | 14.2 | 14.7 | 13.7 | 18.0 | 16.2 |
| 0° | 10° | 92% | 12.9 | 12.8 | 14.3 | 13.1 | 15.9 | 13.9 |
| 0° | 10° | 50% | 13.4 | 13.3 | 14.5 | 13.3 | 16.7 | 14.7 |
| 0° | 5° | 92% | 12.8 | 12.6 | 14.3 | 12.8 | 15.1 | 12.6 |
| 0° | 5° | 50% | 13.6 | 13.4 | 15.3 | 13.8 | 16.9 | 14.4 |
| 45° | 2° | 92% | 12.6 | 12.2 | 14.5 | 12.4 | 14.9 | 11.3 |
| 45° | 2° | 50% | 14.1 | 13.7 | 16.2 | 14.1 | 16.7 | 13.1 |
| 45° | 1° | 92% | 12.5 | 11.6 | 14.4 | 11.2 | 14.6 | 9.2 |
| 45° | 1° | 50% | 13.9 | 13.0 | 15.5 | 12.3 | 16.7 | 11.3 |
| 45° | 0° | 92% | 12.0 | 9.6 | 14.0 | 7.0 | 14.5 | 3.6 |
| 45° | 0° | 50% | 13.2 | 10.8 | 15.4 | 8.4 | 16.8 | 5.9 |
| 90° | 2° | 92% | 12.6 | 12.3 | 14.4 | 12.7 | 15.0 | 12.2 |
| 90° | 2° | 50% | 14.0 | 13.7 | 15.8 | 14.1 | 17.4 | 14.6 |
| 90° | 1° | 92% | 12.6 | 12.2 | 14.3 | 12.3 | 14.5 | 11.0 |
| 90° | 1° | 50% | 13.8 | 13.4 | 16.1 | 14.1 | 16.7 | 13.2 |
| 90° | 0° | 92% | 12.6 | 11.8 | 14.0 | 10.9 | 14.5 | 9.3 |
| 90° | 0° | 50% | 13.8 | 13.0 | 15.7 | 12.6 | 16.6 | 11.4 |
| 180° | 0° | 92% | 12.7 | 12.3 | 14.4 | 12.4 | 14.5 | 11.2 |
| 180° | 0° | 50% | 13.5 | 13.1 | 15.6 | 13.6 | 17.0 | 13.7 |

$K^0$ is the equivalent $K$ *galaxy* magnitude corrected for extinction.
The accuracy on the magnitude limits is $\Delta m \simeq 0.2$.

of sky, and extract the galaxies on the co-added strips, using the algorithms and parameters obtained with the image simulations. We will then iterate on the extraction parameters analyzing the individual strips and comparing to the master list obtained with the co-added strip. In this way, we can optimize the algorithm parameters, no longer being subject to the oversimplified galaxy models used in the image simulations. In addition, this will return us the *selection function*, which we need to know as a function of band, object type and magnitude, star field density, extinction, background (instrumental + sky) and PSF. The accurate determination of the selection function will require a substantial number of strips to be observed a dozen times or so.

To save time, we are considering (following a suggestion by D. Lynden-Bell) focusing on the Galactic Pole, and mimicking the effects of confusion at low galactic latitudes by superposing low galactic latitude fields on top of polar fields, and then attempting to reextract the galaxies that we had indeed detected in the polar field. We would then have to add photon noise to the background to incorporate the effects of extinction and lower surface brightness galaxies at very low galactic latitudes.